# Beyond the RICH: Innovative Photosensitive Gaseous Detectors for new Fields of Applications

P. Carlson[1], C. Iacobeaus[2], T. Francke[1,3], B. Lund-Jensen[1], L. Periale[4,5],
V. Peskov[1], I. Rodionov[6]

[1] Royal Institute of Technology, Stockholm, Sweden
[2] Karolinska Institutet, Stockholm, Sweden
[3] XCounter AB, Danderyd, Sweden
[4] Torino University, Torino, Italy
[5] CERN, Geneva, Switzerland
[6] Reagent Research Center, Moscow, Russia

**Beyond the RICH: Innovative Photosensitive Gaseous Detectors for new Fields of Applications**

P. Carlson[1], C. Iacobeaus[2], T. Francke[1,3], B. Lund-Jensen[1], L. Periale[4,5], V. Peskov[1], I. Rodionov[6]

[1] Royal Institute of Technology, Stockholm, Sweden
[2] Karolinska Institutet, Stockholm, Sweden
[3] XCounter AB, Danderyd, Sweden
[4] Torino University, Torino, Italy
[5] CERN, Geneva, Switzerland
[6] Reagent Research Center, Moscow, Russia

**Abstract**
We have developed and successfully used several innovative designs of detectors with solid photocathodes. The main advantage of these detectors is that rather high gains ($>10^4$) can be achieved in a single multiplication step. This is possible by, for instance, exploiting the secondary electron multiplication and limiting the energy of the steamers by distributed resistivity. The single step approach also allows a very good position resolution to be achieved in some devices: 50 μm "on line" without applying any treatment method (like "center of gravity"). The main focus of our report is new fields of applications for these detectors and the optimization of their designs for such purposes.

**I. Introduction**
The photosensitive wire chamber introduced by J. Seguinot et al [1] and independently by G. Bogomolov et al [2] found immediate application in various fields such as RICH [3], plasma diagnostics [2] and the readout of $BaF_2$ scintillators [4]. The other important step was the developing of gaseous detectors with solid photocathodes, often called gaseous photomultipliers – GPM (see review papers [5,6]). The main advantages of these types of detectors are that their sensitivity could be expanded from VUV to up to visible light by choosing the photocathode material and that they have a very good (< 1ns) time resolution. This allows to enlarge even further the field of application of such detectors to fast RICH, plasma studies and the readout of various fast scintillators. Another boost for the GPM was the invention of micropattern detectors (see [6] and ref. therein). In this work we describe several new applications of the GPM: for the detection of the primary and the secondary scintillation light from noble liquids, for the detection of flames and dangerous vapors and for VUV spectroscopy. The main focus of this paper is on the optimization of the GPM designs, especially micropattern designs, for these particular applications.

**II. Detection of primary and secondary scintillation light from noble gases and liquids with GPMs**
**II.1 Preliminary results**
   There are several projects exploiting excellent scintillating properties of noble liquids (see for example [7] and references therein). Recent examples are the LAr/Xe calorimeter for nTOF [8] and detectors oriented on the search of WIPM [9]. The most common detectors of the scintillation light are still PMTs [10]. As suggested earlier, GPMs with CsI photocathodes can be used to detect the primary scintillation light from noble gases (see [5] and references therein) and noble liquids [11,12]. The advantages of such an approach are the simplicity, low cost, large sensitive area and potentially possibly to operate without a separating window between the noble liquid chamber and the GPM. In preliminary measurements, wire types of GPMs were used [5, 12, 13]. They had reflective CsI photocathodes and were separated from the noble liquid chamber by $CaF_2$ or $MgF_2$ windows [11,12]. Very encouraging preliminary results were obtained with this simple approach. For example primary scintillation light produced by 22 keV X-rays was detected with ~100% efficiency (the solid angle of the detection was about $10^{-2}$). One should note that in the last few years there have been intense developments of various types of GPMs: based not only on wire types of detectors [13], but also on GEM, MICROMEGAS, MSGC and glass capillary plate (CP) [14,6]. So a natural question arises: are the wire types of detectors used in the first measurements the best for detecting primary and secondary scintillation lights, or is there an optimum design of the GPM? To answer this question we performed comparative studies of various types of GPMs for this particular application.

**II.2 Optimization of the GPM for the detection of the scintillation light from the noble liquids**
**II.2-1. Experimental set up**

The set up for comparative studies of various detectors is schematically presented in Fig. 1. The set up is only briefly described here and a detailed description as well as the measurement procedure can be found in [15]. The set up consists of essentially two parts: a scintillation chamber and a gas chamber with test detectors installed inside. The scintillation chamber was flushed with pure noble gases of Ar, Kr and Xe at pressures of p=1-2 atm. The primary scintillation light was produced by radioactive sources (in most cases by alphas). In some measurements the secondary scintillation light produced in the parallel mesh structure was used as well. The scintillation light was monitored by a PMT. Note that the spectra of the primary and secondary emissions of noble gases (see [16]) is very similar to the one of the corresponding liquid, so our set up models very well the liquid scintillation. In most measurements, the gas chamber was separated from the scintillation chamber by a $CaF_2$ window. A few tests were done without the separating window when the scintillating chamber and the detector operated in the same noble gas. The following standard designs of GPM were used for comparative studies: MWPC, PPAC, GEM and MICROMEGAS. Their detailed descriptions are given in [15]. In the case of the MWPC and the PPAC designs with only reflective CsI photocathodes (400 nm thick) were used. In the case of the GEM and MICROMEGAS both designs with reflective and semitransparent (20 nm thick) photcathodes were tested. In addition to this, we also tested two innovative designs of GPMs: a CP and a photosensitive RPC. CPs with capillary hole diameters ranging from 12 to 700 μm were studied. In contrast to our previous work, in these studies only capillaries with reflective photocathodes were tested: with the cathodes coated by a CsI layer and with both cathodes and inner walls of capillaries coated by CsI [6]. The design of the RPC was similar to the one described in [17]. The 5 cm in diameter anode disc was made of Pestov glass with a resistivity of $10^{10}$ $\Omega$cm. The cathode was a metallized (~10 nm of Cr) $CaF_2$ disc (the entrance window) coated with a semitransparent 20 nm thick CsI layer. The gap between the electrodes was 0.5 mm. All detectors were tested in a mixture of Ar or He with various quenchers at p=1 atm. There was also a possibility of introducing a $^{55}$Fe radiation inside the drift gap (in the case of PPAC and RPC- directly inside the detector volume) of any of the above described detectors. From the ratio of the amplitude of the signal produced by the $^{55}$Fe and the signal due to the scintillation light, one could estimate the number of detected photoelectrons [6,15]. To estimate the QE, measurements in an ionization chamber mode were done (see [15] for more details) with a Hg lamp used as a light source. This allowed the data to be normalized to the same QE so that a correct comparison between various detectors could be done. In the case of the measurements without the separating window, tests were done with GEMs and CPs (100 μm holes diameter) with only their cathodes coated by a CsI layer.

**II.2-2 Results:**
**1I.2-2a) Detectors with windows:**

Our main results obtained with various detectors in He-based mixtures are summarized in Table 1. In some cases ethylferrocene vapors (EF) were added to these mixtures. The He based mixtures allow 2-5 times higher gains to be achieved compared to Ar -based mixtures. Note that in Table1 only data for CPs with 100 μm hole diameters are included. This geometry offer the maximum gas gains (see Fig. 2). In our experiments the QE of the CsI photocathodes varied from detector to detector. The measurements in the ionization chamber mode however, (measurements for each particular detector of the photocurrent produced by the Hg lamp) allowed necessary corrections to be made (see the correction factor η in Table.1, which includes current, geometrical and CsI coating area corrections) and thus all detectors could be compared in such a way as if they had the same QE. From the data presented in Table. 1, one can see that among the detectors with reflective photocathodes, the highest gains and quantum efficiency ($QE_r$) was achieved with the PPAC. Wire-type detectors offer less gains and QE, however in some experimental environments they can still be very attractive since they do not have any sparking problems.  Among detectors with semitransparent photocathodes, the highest gains and quantum efficiency ($QE_s$) was obtained for the RPC. One should point out the very attractive feature of this innovative photosensitive detector: RPCs with CsI photocathodes can operate at gains of up to $10^6$ without any sparking problems. As was shown in [17], the RPCs also have excellent time resolutions of much better than 1 ns. We will in the next section describe another unique property of this detector: excellent position resolution ~50 μm.  The explanation to the question of why the PPAC and the RPC have the highest QE is given in [15]. It is based on the fact that both detectors operate at high electric fields (>30kV/cm) between the electrodes (and the QE is dependant on the electric field [5,18] as well as having a 100% photoelectron detection efficiency. The other types of GPMs either have a lower mean electric field near the cathode or, due to their geometrical features do not have a full photoelectron collection efficiency.

**II.2-2b) Windowless detectors**

The operation of the CPs and the GEM with reflective (400 nm thick) and semitransparent (20 nm thick) CsI photocathodes was studied in pure Ar, Kr and Xe at p=1-2 atm. For gain measurements an alpha source was used. It allows obtaining gain vs. voltage characteristics. Then the souse was removed and the voltage was increase till breakdowns appeared. From the known breakdown voltage and gain vs. characteristics a maximum achievable gains were estimated. Both detectors could operate in noble gases only at relatively low gains <50 (see Table. 2). Even at these gains some instability in time was observed with the CPs. More work should be done to understand the nature of this instability. For the time being, however, we conclude that GEMs exhibit a better performance. Unfortunately, in order to reach gains of >$10^3$ necessary for the single electron detection, two or more GEMs operating in tandem mode are required [9,11], and this causes additional experimental complications.

### III. GPMs for spectroscopic studies and high position resolution measurements in the VUV region

GPMs with solid (CuI) photocathodes were successfully used in the past in plasma research for imaging and spectral measurements [19]. Recently we have developed (a description follows below) GPMs oriented on similar applications but having much better characteristics than in the version described in [19].

### III.1 Detector design and experimental set up

The detector is actually a modified version of the photosensitive RPC described above (see Fig. 3). The cathode was manufactured from a 50x50 mm$^2$ GaAs wafer 0.5 mm thick, with a resistivity of $10^6$ Ωcm. The inner part of the cathode was coated by a 0.4 nm thick CsI layer. The anode was manufactured from 50x50 mm$^2$ Pestov glass with a resistivity of $10^{10}$ Ωcm. The inner part of the anode was covered by Cr strips of 50 μm pitch. Each of the 20 central strips was connected to individual charge sensitive amplifiers. The other strips were connected together to one common charge sensitive amplifier. The gap between the cathode and the anode planes was 0.5 mm. The detector worked in a mixture of Ar with 10% ethane at p=1 atm. For position resolution measurements a screen was used with a 50 μm wide slit oriented perpendicular to the electrode's plane. The slit was irradiated with a Hg lamp placed 50 cm away. The slit could be moved in a direction perpendicular to the strips with micron accuracy. In the case of the two-dimensional measurements, the detector itself could be moved in the direction perpendicular the electrode's plane with micron accuracy (scanning mode). For the QE measurements the RPC was operated in ionization chamber mode. The photocurrent (produced by the light from the Hg lamp light passed through a 192 nm narrow band filter) was measured by a Ketley picoampermeter. The QE was estimated by comparison of this photocurrent with the one measured from the calibrated Hamamatsu photodiodes R1187 and R414.

### III. 2 Results

Fig. 4 shows the results of the position resolution measurements performed with the moving slit. When the slit was aligned exactly along one strip, the maximum counting rate was measured from this particular strip. When the slit was aligned in between two strips, the highest counting rates were measured from these strips. This picture was periodically repeated during the slit movement. One can conclude from this data that a position resolution of about 50 μm was achieved. Note that this very good position resolution was achieved in simple counting (digital) mode without applying any treatment method (like "center of gravity"). In contrast to traditional RPCs, which have very limited rate characteristics (usually $< 10^4$ Hz/cm2) [20]), our detector, due to the low resistivity of the cathode, could operate at counting rates of up to $10^6$ Hz/cm2 without any sparking or charge build-up up effect (see Fig. 5). The QE of this detector depends on the angle φ at which the Hg light hit the CsI cathode: it was 15% at φ=30° and 2.7% at φ=5°.

**IV. GPMs for the detection of UV flames, sparks and dangerous vapors**

In some applications it is necessary to detect sparks or flames in daylight conditions. In this work we investigated the possibility of using GPMs for this purpose. The detectors used in this study were based on commercially available sealed single wire counters [21]. They are made of 5 cm long stainless steel tubes with diameter of 15 mm, having, depending on the design, 0.1 or 1 mm diameter of anode wires. The tubes have glass interface ( a glass supporting structure)at both ends between the anode wire and the cathode. All the detectors have a $LiF_2$ window. For photocathode manufacturing the glass seal at the interface and the window were opened. The CsI and CuI photocathodes were manufactured by usual technique [22]. The detectors were then pumped to a vacuum of $10^{-6}$ Torr, heated to 50°C, filled with gas (P10 with ethylferrocene vapors at total pressure of 1 atm, see [22] for more details) and sealed. We also made the first attempts to develop CsTe photocathodes (see [6]). Fig. 6 shows the QE of our detectors as well as the typical emission spectra of the flames in air. One can see that the QEs of CuI, CsI and especially of CsTe photocathodes, overlap with the flame emission spectra. At the same time, such detectors are practically insensitive to visible light. Results of some of our tests are presented in Table 3. One can see that a flame from a cigarette lighter could be reliably detected at a distance of 6-7 m. Analysis of this data shows that the efficiency of the detection and the signal to noise ratio was comparable to that of commercially available one. A modified version of the device (see Fig.7) can also detect smoke and dangerous vapors. It contains a compact VUV lamp (a corona discharge in Ar or Xe) and several single wire GPMs surrounding the lamp at a distance of a few mm. Any changes in the transmission of the gap between the VUV lamp and the GPM causes a change in response in the detector which is used to identify smoke or dangerous levels of various gases. Preliminary experiments show that ~1% of the vapor concentration in air can be reliably detected.

**IV. Conclusions**

We have developed and successfully tested several innovative designs of photosensitive detectors with solid photocathodes, for example:

•CPs with the cathodes and the inner walls coated with a CsI layer;

• RPCs with CsI photocathodes;

- gaseous detectors with CsTe photocathodes.

The main feature of these detectors is that rather high gains (>$10^4$) can be achieved in a single multiplication step. The single step approach also allows a very good position resolution to be achieved in some devices: 50 μm "on line" without applying any treatment method (like center of gravity). These detectors were successfully used for new applications:

1) detection of primary and secondary scintillation lights from noble gases and liquids;
2) as a linear detector matrix in the focal plane of vacuum spectrograph,
3) for a scanner of VUV images,
4) detection of VUV radiation from various flames,
5) detection of dangerous vapours in air.

We believe that some of these detectors (for example photosensitive RPCs) or their combinations could be useful for the RICH application.

**Figures caption:**
Fig.1 Experimental set up for comparative studied of various GPMs
Fig.2 Maximum achievable gains for CPs with various capillary hole diameters: measurements were done in He+3%CH4+EF=1atm. Radioactive source-$^{55}$Fe.
Filled squares-data for CPs with inner coating of capillaries with CsI [5].
Fig. 3 A schematic drawing of high position resolution photosensitive RPC
Fig. 4 Number of counts from various strips measured at three different positions of the collimated slit.
Fig.5 Gain vs. counting rate characteristics for metallic PPAC (filled squares) and for photosensitive RPC (triangles and diamonds represent RPC's behavior at two different gains)
Fig.6 Typical flame spectra in air (open symbols) [23] and QE of single wire detectors with various photocathodes: CsI (diamonds), CuI with EF adsorbed layer (large squares), CsTe (small squares)
Fig.7 A schematic view of the device for detecting dangerous vapors in air

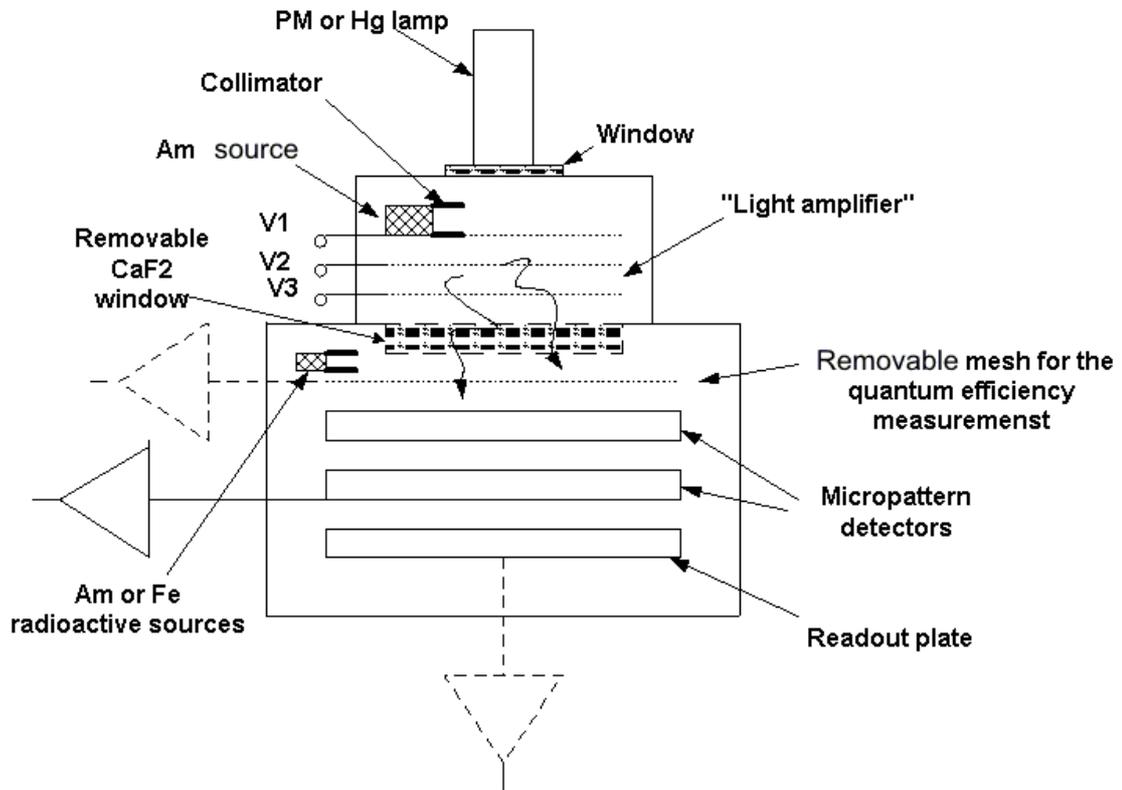

Fig. 1

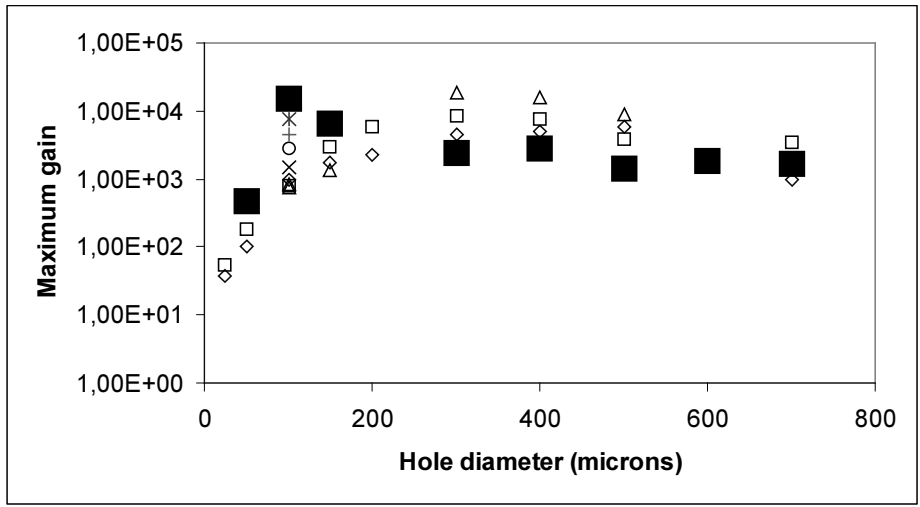

Fig. 2

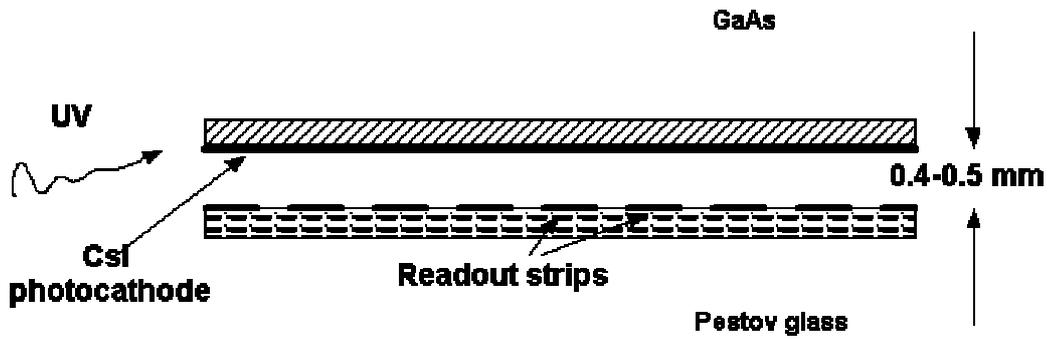

Fig. 3

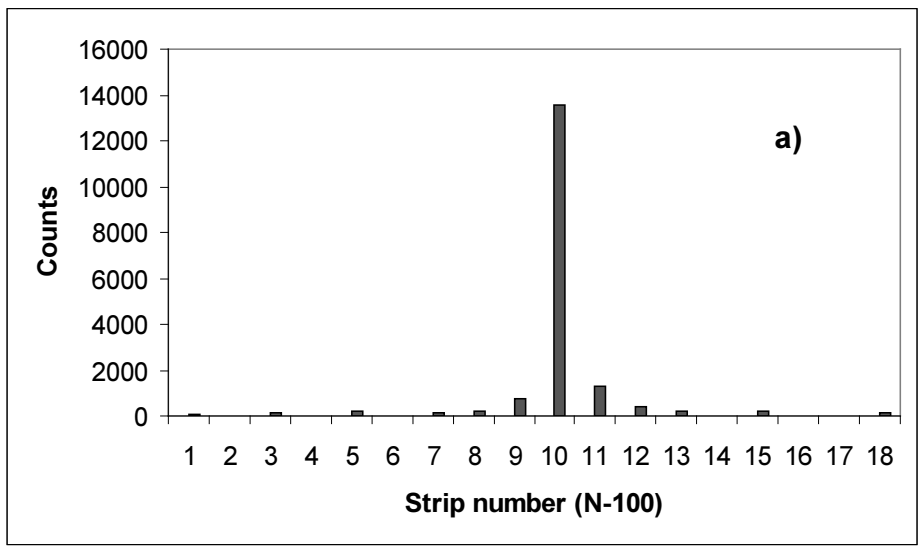

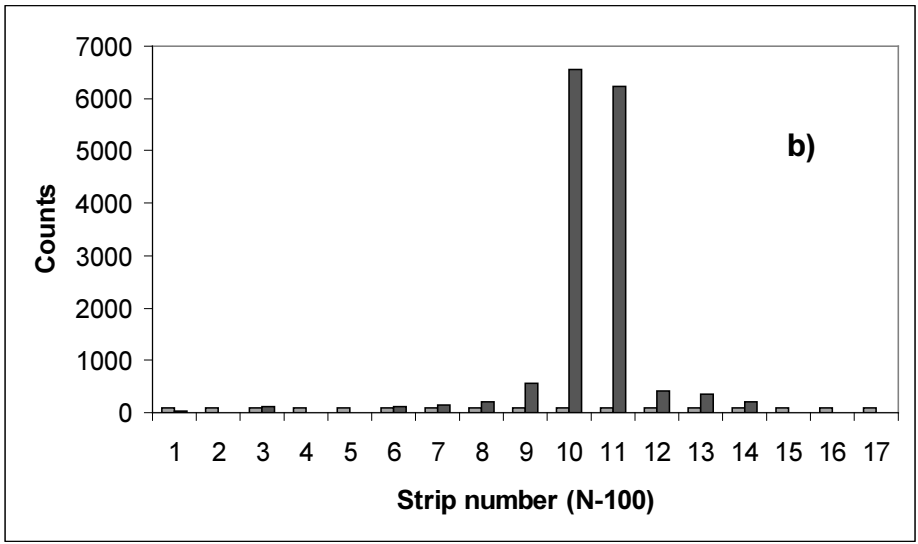

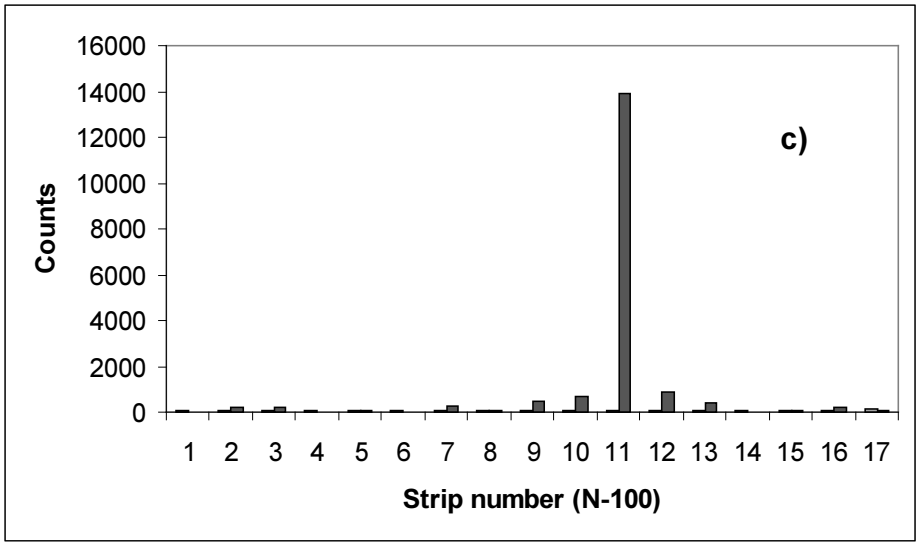

Fig. 4

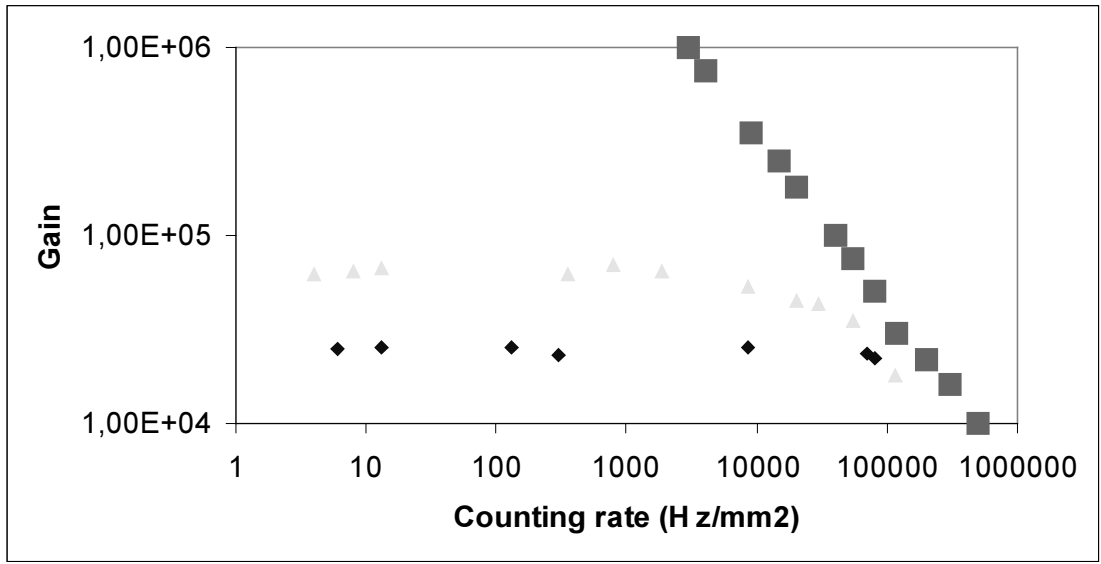

Fig. 5

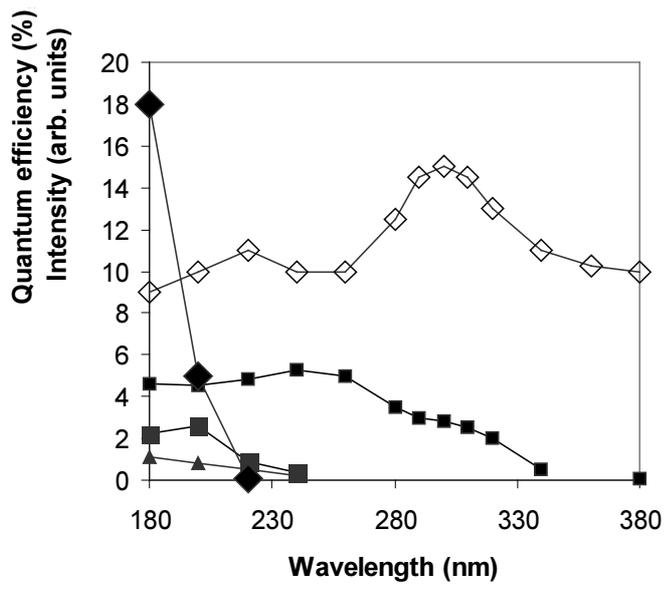

Fig. 6

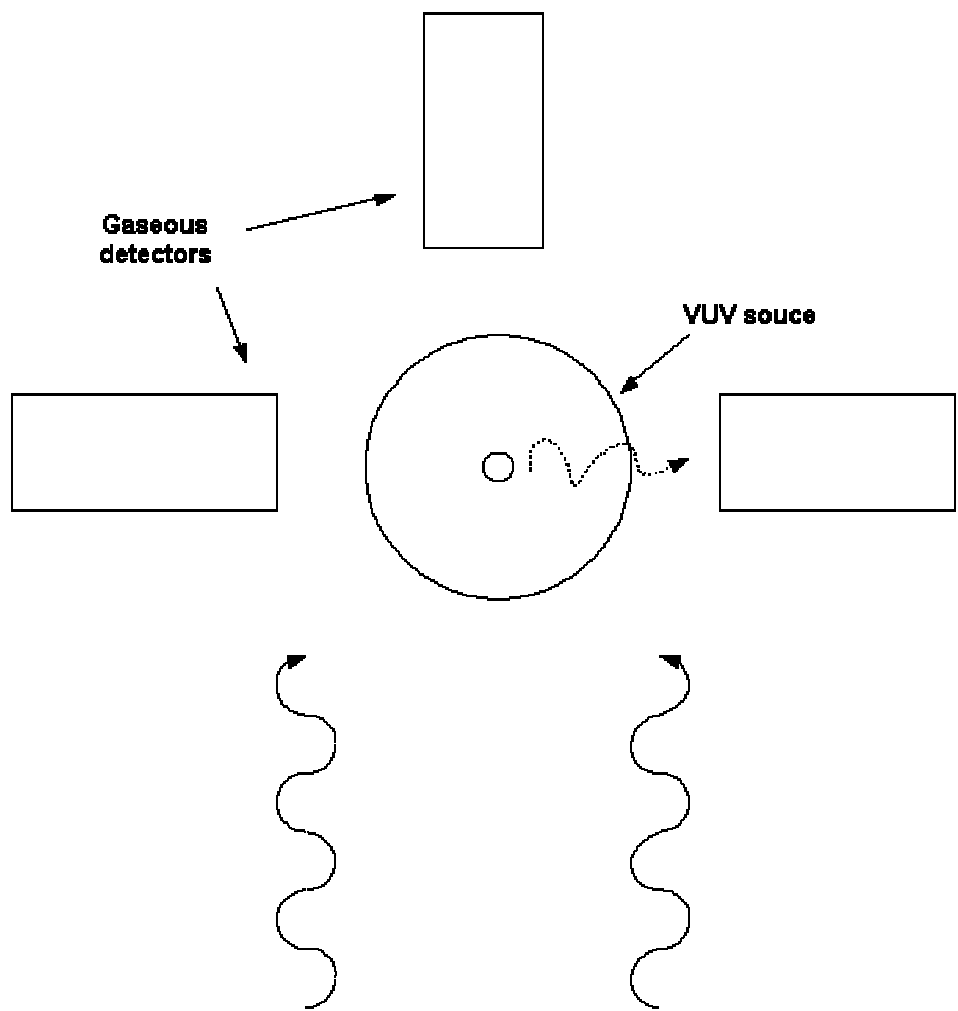

Fig. 7

## Table-1

### Efficiency ( to Xe scintillation light) of various window –type GPMs

| GPM type | Gas | Safe gain | QEs (%), {η} | QEr (%), {η} |
|---|---|---|---|---|
| MWPC | He+20%CH4+EF | $3 \times 10^4$ | 4.3 {1.3}] | 11.2 {1.25} |
| PPAC | He+ 1%CH4+EF | $10^5$ | 7.4 {2.1} | 19.6 {1} |
| RPC | He+ 1%CH4+EF | $10^5$-$5 \times 10^5$ | 3.2 {2.2} | |
| MICROMEGAS | Ar+5%CH4 | $10^4$ | 4.5 {2.4} | |
| GEM | He+10%CO2+EF | $\sim 10^3$ | 1.9 {2.6} | 3.1 {1.6} |
| CP (CAT mode) | He +3%isob+EF | $< 3 10^3$ | 2.2 {2.7} | 2.9 {2} |

## Table-2

### Efficiency of windowless detectors with reflective CsI photocathodes operating in pure noble gases

| Detector | QEr (%) in Xe, {η} | Gain | QEr (%) in Kr, {η} | Gain | QEr (%) in Ar, {η} | Gain |
|---|---|---|---|---|---|---|
| GEM | 1.9  {1.8} | ~30 | 1.4 {2} | ~20 | 3.6 | ~25 {1.6} |
| CP | 1.5   {2.8} | ~25 | | | | ~20  {2.8} |

**Table-3**

**Counting rate (N) obtained from single wire detectors with various photocathodes (PC)**

| PC \ N | Dark | Daylight | Match 1m | Cigarette Lighter 7m |
|---|---|---|---|---|
| CsI (Flushed by gas) | 2 | 5-10 | 920 | 112 |
| CuI (flushed by gas) | 2 | 3 | 24 | 7 |
| CsI (sealed) | 2 | 3-5 | 431 | 65 |
| CuI (sealed) | 1 | 3 | 11 | 5 |
| CsTe (sealed) | 5 | 23 | 5560 | 624 |